\documentclass[twocolumn,amssymb,aps,prb]{revtex4-2}

\usepackage{dcolumn}
\usepackage{xfrac}
\usepackage[utf8]{inputenc}
\usepackage{graphicx}

\usepackage{graphics,epsfig,graphicx}
\usepackage{color}
\usepackage{xcolor}
\usepackage{cancel}
\usepackage{makeidx}
\usepackage{latexsym}
\usepackage{fancyref}
\usepackage[unicode=true, bookmarks=true, bookmarksnumbered=false, bookmarksopen=false, breaklinks=false, pdfborder={0 0 1}, backref=false, colorlinks=true]{hyperref}
\usepackage{cleveref}

\usepackage[normalem]{ulem} 
\usepackage{color} 

\begin{document}

\title{Enhancement of optical coherence in $^{167}$Er:Y$_2$SiO$_5$ crystal at millikelvin temperatures}

\author{N.~Kukharchyk}
\affiliation{Experimentalphysik, Universit\"at des Saarlandes, D-66123 Saarbr\"{u}cken, Germany}
\email[E-mail: ]{nadezhda.kukharchyk@wmi.badw.de}

\author{D.~Sholokhov}
\affiliation{Experimentalphysik, Universit\"at des Saarlandes, D-66123 Saarbr\"{u}cken, Germany}

\author{A.~A.~Kalachev}
\affiliation{FRC Kazan Scientific Center of RAS, 420029 Kazan, Russian Federation}

\author{P.~A.~Bushev}
\affiliation{Experimentalphysik, Universit\"at des Saarlandes, D-66123 Saarbr\"{u}cken, Germany}
\affiliation{JARA-Institute for Quantum Information (PGI-11), Forschungszentrum J\"{u}lich, 52428 J\"{u}lich, Germany}

\date{\today}

\begin{abstract}
	
Er$^{3+}$:Y$_2$SiO$_5$ crystal is a promising candidate with a great variety of its potential applications in quantum information processing and quantum communications ranging from optical/microwave quantum memories to circuit QED and microwave-to-optics frequency converters. Some of the above listed applications require ultra-low temperature environment, i.e., temperatures $T\lesssim0.1~$K. Most of the experiments with erbium doped crystals have been so far carried out at temperatures above 1.5~K. Therefore, only little information is known about Er$^{3+}$:Y$_2$SiO$_5$ coherence properties at millikelvins. Here, we investigate optical decoherence of $^{167}$Er:Y$_2$SiO$_5$ crystal by performing 2- and 3-pulse echo experiments at millikelvin temperature range and at weak and moderate magnetic fields. We show that the deep freezing of the crystal results in an increase of optical coherence time by one order of magnitude compared to temperature of 1.5 Kelvin and magnetic field of $\sim$0.2~T, taken as a reference point. We further describe the detailed investigation of the decoherence mechanisms in this regime.
\end{abstract}
\maketitle
\section{Introduction}

Rare-earth (RE) doped solids have gained considerable interest from quantum communication~\cite{GisinRMP2011} and quantum information processing community~\cite{Thiel2011} due to exceptionally long coherence time of their spin~\cite{Sellars2015,Sellars2018} and optical degrees of freedom~\cite{Boettger2009}. 
Crystals doped with Kramers RE ions (those with odd number of electrons), such as Nd$^{3+}$, Yb$^{3+}$ and Er$^{3+}$, are studied in a view of different applications such as
(a) optical quantum memories, due to the presence of optical transitions inside telecommunication bands~\cite{Lauritzen2010, Gisin2014, Tittel2015, Boettger2016, Goldner2016, Faraon2017}; (b) efficient microwave quantum memories, because of long coherence time of electronic and nuclear spins~\cite{Probst2015, Morton2015, Morton2018}; (c) circuit QED, due to large g-factor~\cite{Probst2013,Longdell2016,Saito2018}; (e) microwave-to-optical frequency converters, due to the addressable transitions is optics, microwave and RF frequencies and large g-factor~\cite{Longdell2018_2,Thiel2018}.
Nevertheless, the major challenge of working with Kramers ions is associated with their large unquenched electron magnetic moments, which exhibit rapid decoherence as a result of the increased coupling to phonons (process known as spin-lattice relaxation)~\cite{Afzelius2008, Afzelius2017} and to other spins via magnetic dipolar interactions (spectral and instantaneous diffusion, spin-spin relaxation)~\cite{Boettger2006, Morton2018}.

Numerous methods allow to lessen the influence of the decoherence processes. 
One of such methods relies on using the zero first-order Zeeman (ZEFOZ) shift technique, where an optical photon is mapped to a transition which is insensitive to the magnetic field fluctuations~\cite{McAuslan2012}. 
In some substrates, e.g., low-symmetry Y$_2$SiO$_5$ (YSO) crystal, such ZEFOZ transitions may occur at zero magnetic field~\cite{Longdell2018} or at specific magnitudes and directions of the applied magnetic field~\cite{Afzelius2018}. 
Besides, YSO is magnetically quite crystal where rates and amplitudes of fluctuating fields induced by present nuclear spins are weak which makes it easier to achieve higher coherence times.
As a result, the electronic spin coherence time may reach milliseconds timescale. 
Another way is to freeze electronic spin bath by applying a strong magnetic field of 7~T at a relatively low temperature of 1.5~K and, e.g., to write coherent optical pulses into hyperfine spin states of a Kramers ion. By using this prescription a record-long $T_2\simeq1.3~$s has been recently obtained for $^{167}$Er$^{3+}$ ions~\cite{Sellars2018}. 
One more promising way involves the possibility to map optical photons into nuclear spins of a crystal host, such as Y$^{3+}$ in the case of Y$_2$SiO$_5$ (YSO) crystal~\cite{Thierry2018}.


In this article, we present an experimental investigation of optical coherence of $^{167}$Er:Y$_2$SiO$_5$ (Er:YSO) crystal at millikelvin temperatures. 
This thermal range is very attractive due to few possible applications, which are hardly accessible at conventional temperatures above 1.5~K: for instance, the direct interface between superconducting qubits and RE optical or spin degrees of freedom with a view of the application in microwave quantum memory~\cite{Kubo2011}, and microwave-to-optical frequency converters~\cite{Obrien2013,Longdell2018_2}. 
At ultra-low temperatures, it is possible to attain nearly full polarization of the electronic spin bath, which in turn quenches the major sources of optical/spin decoherence, i.e. direct and indirect flip-flops of surrounding electronic spins~\cite{Probst2015}. 

In solid state optical spectroscopy, the millikelvin temperature range is very challenging to work at due to low thermal conductivity of substrate at these temperatures and appearance of well-known effect of Kapitza resistance which appears on the interface of dielectric material and metal. These lead to higher effective temperature of the spin system and substrate in comparison to the temperature of the cryostat~\cite{Runge2003}.  This effective temperature also differs between the spectroscopy and echo experiments as a result of different excitation energies applied in these experiments~\cite{Kukharchyk2017}. In addition, the energy of a single phonon induced by the direct process is comparable to thermal energy, while thermal conductance properties of substrate and interface are getting strongly reduced below 1~K~\cite{Pobell2007}.

In our recent work, we have demonstrated the increase of optical coherence time of isotopically purified $^{166}$Er:$^7$LiYF$_4$ at $T<1$~K~\cite{Kukharchyk2017} by nearly two orders of magnitude while cooling down below 1.5 K at moderate fields. The LiYF$_4$ is however a challenging substrate: the Fluorine possesses large nuclear spin moment, which is a source of magnetic noise and strongly limits the electronic coherence of Erbium even below 1~K. 
In the following, we discuss the decoherence sources in the magnetically quite Er:YSO at the magnetic fields up to 300~mT and below 1~K.

\section{Experimental setup}
\begin{figure}[t!]
	
	\includegraphics[width=1\columnwidth]{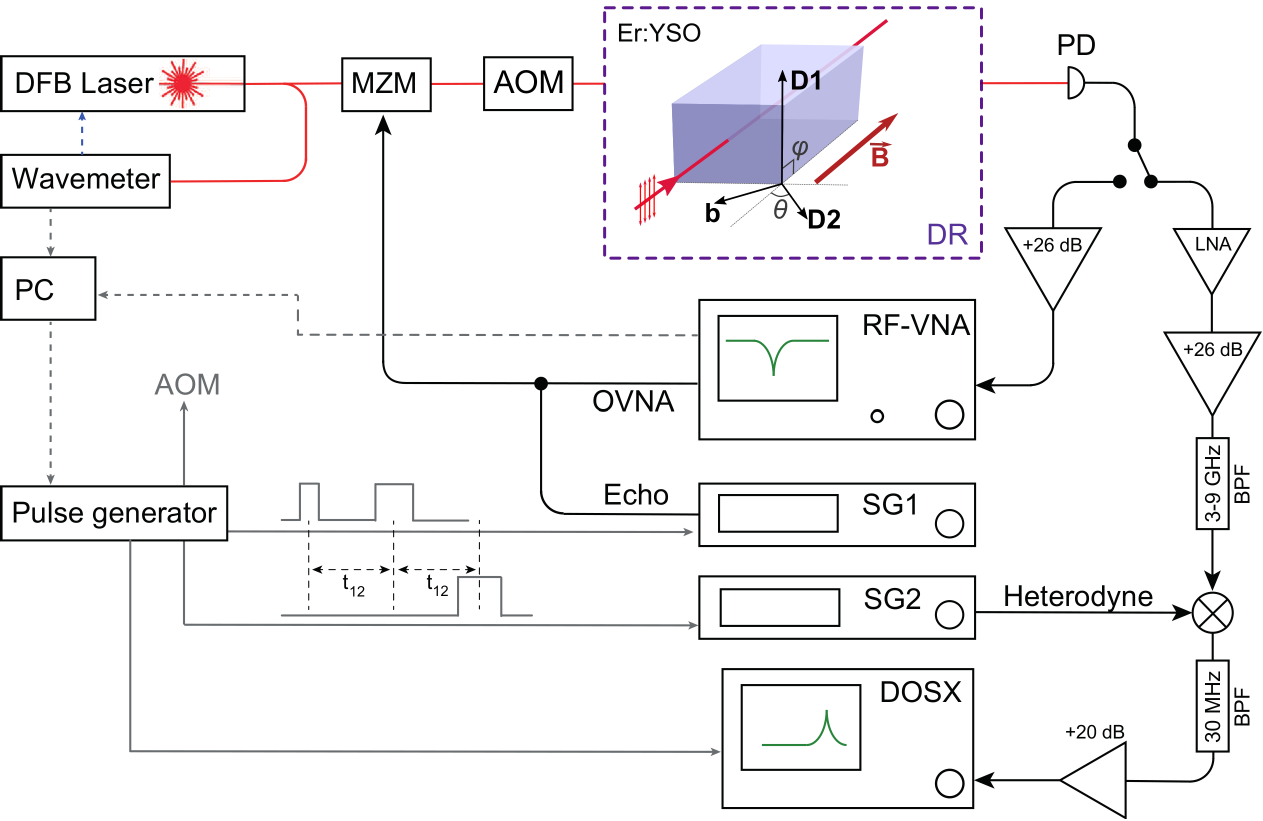}
	
	\caption{(Color online) Sketch of the experimental setup. 
		OVNA stands for the optical vector network analysis and for heterodyne echo detection. MZM stands for the Mach-Zehnder intensity modulator. AOM is the acousto-optical modulator. DR is the dilution refrigerator. PD is the high-speed InGaAs photodetector. SG1 and SG2 are signal sources for the generation of echo sequence and heterodyne detection, respectively. RF-VNA is the radio-frequency vector network analyser. AOM and signal generators are triggered by using pulse generator. DOSX is the digitizing oscilloscope. The experiment is controlled by using PC. The sketch of the crystal shows the orientation of the optical symmetry axes with respect to the laser and magnetic fields.}
	\label{Setup}
\end{figure}

We investigate an Er:YSO  single crystal doped with 0.005\% atomic concentration of $^{167}$Er$^{3+}$ ions grown by Scientific Materials (Bozen, USA). The crystal has dimensions of 3 x 4 x 6~mm and its faces are AR coated for 1539 nm wavelength transmission. 
The YSO is a low symmetry crystal where Erbium ions replace Yttrium ions leading to the presence of two symmetry classes (site 1 and site 2) and two magnetic classes of Erbium ions~\cite{Sun2008}.
In the experiment presented, the orientation of the crystal in the magnetic field ($\theta=45^{\circ}$ and $\varphi=90^{\circ}$, see Fig.~\ref{Setup}) allows for lifting off the magnetic class degeneracy~\cite{Probst2013,Probst2015}. The optical pulses are propagating along the magnetic field, and polarization of the light is set along $D_1$ axis of the crystal. The crystal is placed inside a copper sample holder which is thermally anchored to the mixing chamber of the optical dilution refrigerator BF-LD-250 with calibrated cooling power of $450~\mu$W at $T=0.1~$K, see refs.~\cite{Probst2015,Kukharchyk2017} for details.

A schematics of the experimental setup is outlined in Fig.~\ref{Setup}. The erbium-doped free-running fiber laser (NKT Photonics Adjustik E15) emits a continuous signal at the frequency which is about $6~$GHz below the observed $I_{15/2} \leftrightarrow I_{13/2}$ optical transition for the site 2 at zero field ($\omega_0/2\pi = 194802~$GHz).  The laser frequency is stabilized by using an optical wavelengthmeter (High-Finesse WS6-200). Optical vector network analysis (OVNA) is used for the transmission optical spectroscopy of the sample~\cite{Kukharchyk2018,Kukharchyk2020}. For that purpose the signal generated by RF vector network analyzer (RF-VNA) is used to create optical sidebands via Mach-Zehnder intensity modulator (MZM). The total power of the laser beam is adjusted by acousto-optical modulator (AOM) and is set to $30~\mu$W during spectroscopy experiments and to $5~$mW for photon echo experiments. The light beam is focused into the sample yielding an estimated beam waist of $100~\mu$m. The transmitted signal is detected by a high-speed InGaAs photodetector (PD). The intereference between optical sidebands and the carrier signal produces microwave signal at the excitation frequency. This signal is amplified, and its magnitude, group delay and phase are measured by RF-VNA. 

Pulsed optical spectroscopy (2PE and 3PE) is implemented by using heterodyne method where the echo sequence is created by modulating the laser carrier via MZM. The modulating signals are powered by the RF signal generator SG1. Microwave pulses of varied length and durations create two optical sidebands (red- and blue-detuned), one of which (blue-detuned) is resonant with an erbium transition. The carrier and the red-detuned sideband are detuned from the erbium transitions. Another triggering sequence opens AOM only for the time of the echo generation pulses and its detection. The light power in the interacting sideband is estimated to be about $1~$mW. The interference between the carrier pulse and the echo signal is detected by a high-speed photoreceiver. The detected microwave pulse is amplified by using low-noise amplifier (LNA) and conventional $+26~$dB amplifier. The amplified microwave signal is mixed down to $30~$MHz with the help of the heterodyne SG2. After the band-pass filtration, the echo signal is finally detected by using a digitizing oscilloscope DSO-X. The full-experiment is controlled by using PC and MATLAB scripts.

\section{Spectroscopy}
\begin{figure}[t!]
	\includegraphics[width=1\columnwidth]{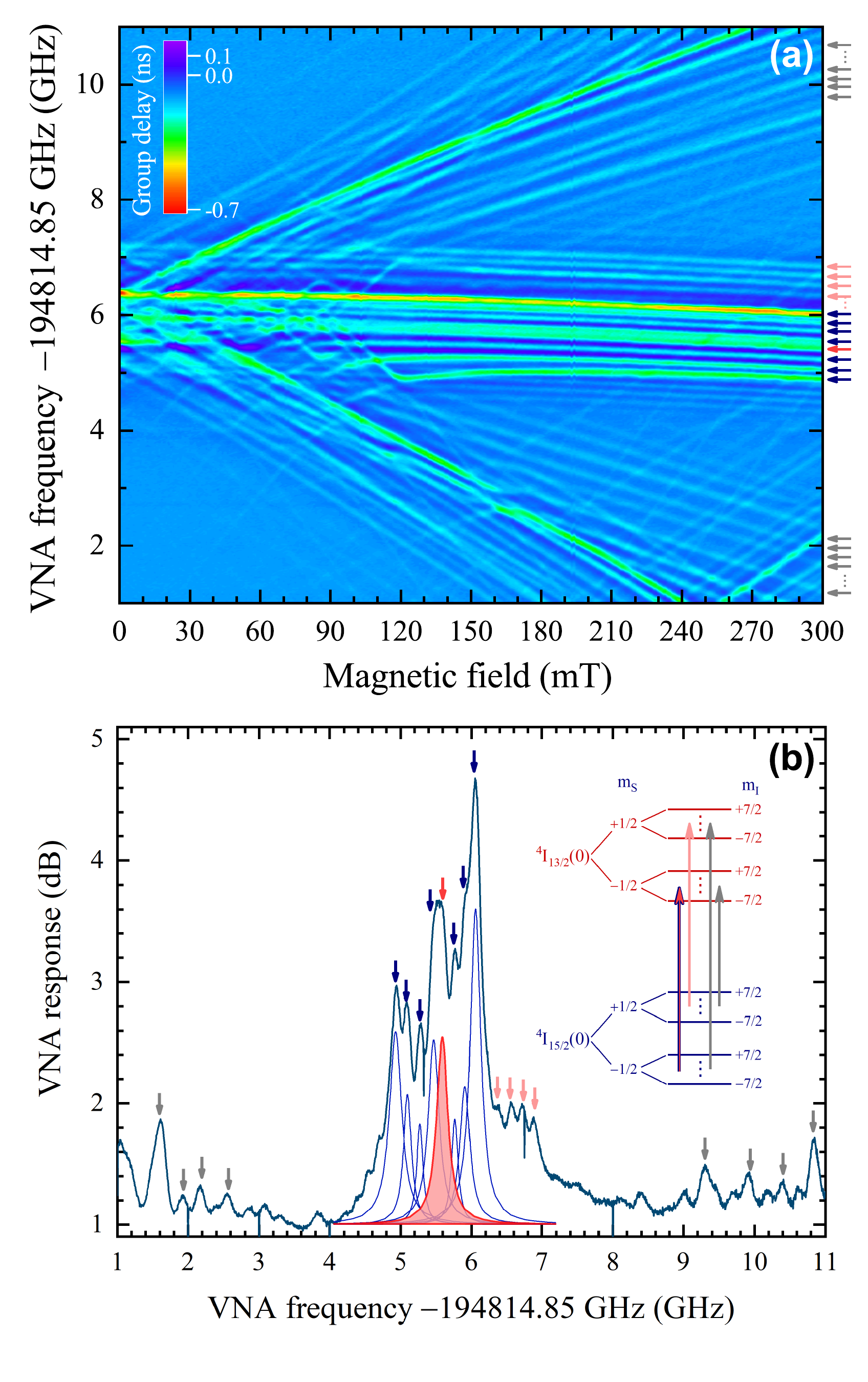}
	\caption{(Color online) (a) The OVNA group delay transmission spectrum of Er:YSO sample measured at $T=12~$mK as a function of the applied magnetic field. 
		(b) The OVNA amplitude transmission spectrum measured at $B=280~$mT.
	The inset depicts level scheme of ${}^{167}$Er. Colour-coded arrows indicate transitions between Zeeman levels with the same nuclear projection $m_I$. Using same colour-code, arrows indicate corresponding transitions to the right of colour-plot in (a) and on top of the absorption spectrum in (b). 
	The red arrow shows the transition which is used to study the optical coherence.}
	\label{Spectrum}
	\end{figure}

Figure~\ref{Spectrum}(a) demonstrates the group delay transmission spectrum  measured at the base temperature of the dilution refrigerator $T=12~$mK as a function of the applied magnetic field. At weak fields, the magnitude of the transmitted signal $\vert S_{21}(\Omega)\vert$ consists of many overlapping lines, and it is therefore difficult to assign each line to a particular optical transition. However, the measured group delay of the microwave signal yields much cleaner spectrum, which is also insensitive to the intensity fluctuations. The group delay signal $\delta \tau = \partial \arg \big(S_{21}(\Omega)\big) /\partial \Omega$  is automatically computed by RF-VNA. 

The transmission spectrum consists of four groups of lines with large (grey arrows) and small (red and blue arrows) optical g-factors, see Fig.~\ref{Spectrum}. All groups can be identified at magnetic field above 0.2 Tesla, where the electron Zeeman term prevails and the electronic spin becomes a good quantum number. 
The two groups of lines with large g-factors correspond to the transitions between the levels with opposite projections of the electronic spin. 
The two other, central, groups correspond to the transitions between the levels with same projection of electronic spin. The level-scheme with colour-coded transitions is presented in the inset of Fig.\ref{Spectrum}~(b), respectively coloured arrows are pointing to the corresponding transition lines in the absorption spectra in Fig.~\ref{Spectrum}.

We focus our study of optical coherence on the central group of lines with small optical g-factor, marked with blue arrows in Fig.~\ref{Spectrum}a. There are 8 resolved lines in this multiplet, which are attributed to the optical transitions between Zeeman states with the same projection of electronic spin $m_S=-1/2$, but different nuclear spin projection $m_I=-7/2...+7/2$ while $\Delta m_I = 0$. 
From the spectra, we derive the effective optical g-factor of the central multiplet to be equal $g_{opt}\simeq 0.2$. Taking advantage of the available g-tensors of the ground and excited states~\cite{Thiel2010}, we simulate the g-factor of the ground state to be equal $g_{spin}\simeq 1.7$ and g-factor of the surrounding site 1 spin to be equal $g_{env}\simeq 4$. The g-factor values vary with the magnetic field, and therefore $g_{opt}$, $g_{spin}$ and $g_{env}$ are effective g-factors allowing to describe the magnetic field range ($100\,\textendash\,300$)\,mT.

The optical transmission spectrum, i.e. the magnitude of the RF-VNA response, measured at $B=280~$mT is shown in Fig.\ref{Spectrum}(b). Optical density of the observed transitions can be calculated from the dB-signal by using the simple theoretical OVNA model for the central multiplet $\alpha L \sim (0.5-1)$~\cite{Kukharchyk2018}. The full-width of the optical inhomogeneous broadening is relatively small $\Gamma_{opt}\simeq (190\pm 20)$~MHz, fits of spectral lines are shown in Fig.~\ref{Spectrum}(b).

\section{Optical echo measurements}

\begin{figure}[t!]
	\includegraphics[width=0.8\columnwidth]{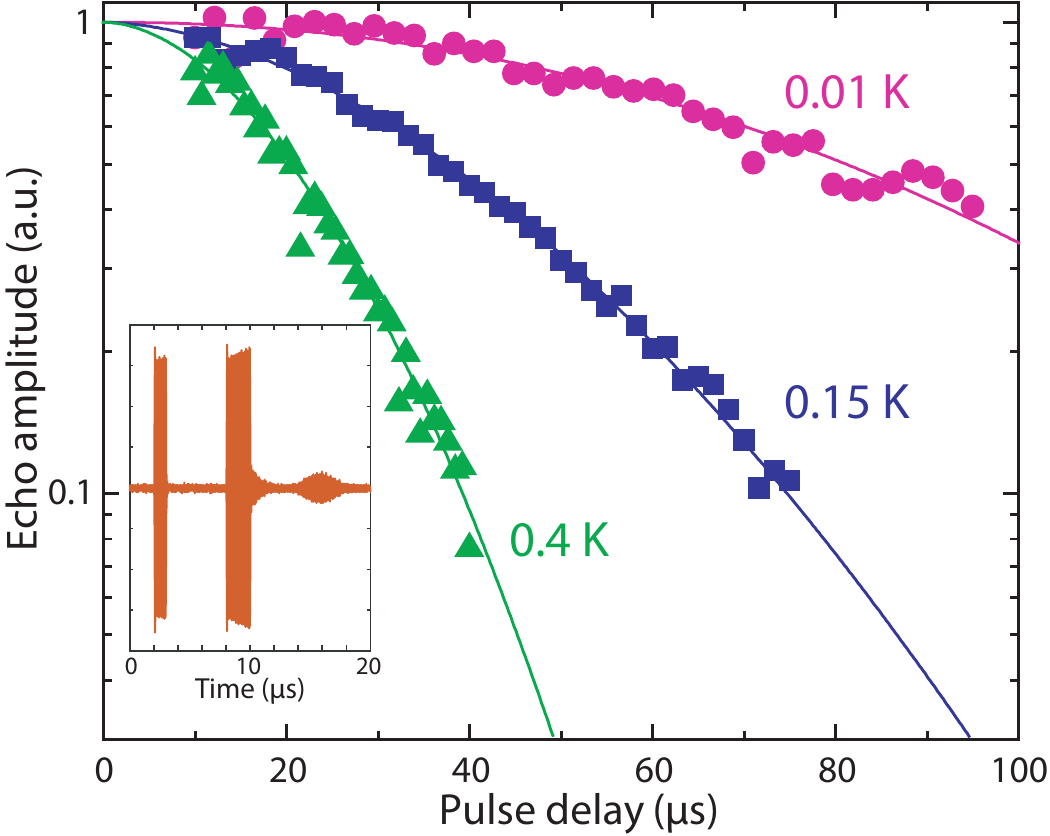}
	\caption{(Color online) Two-pulse echo decay measured at 280 mT at temperatures of 0.012~K, 0.15~K and 0.4~K. The solid line  represents the fit of the data.
	In the inset, the heterodyne 2PE sequence and the echo envelope are shown for the delay of 7~$\mu$s.
	}
	\label{Echo_decay}
\end{figure}

Coherent optical spectroscopy of the present sample is carried out by two- and three-pulse echo (2PE and 3PE) experiments. The 8 lines of the central multiplet demonstrate similar optical coherence properties, therefore, in the following we discuss results measured for the line in the middle, which is marked by the red arrow, see Fig.~\ref{Spectrum}(b). This line corresponds to the transition between Zeeman states with nuclear magnetic number $m_I=1/2$.
The length of the $\pi/2$-pulse is optimized to yield the maximum echo amplitude and is equal to 1.2$~\mu$s which corresponds to the Rabi frequency of $\simeq2\pi \cdot 200$~kHz. An example of the heterodyned echo sequence for the pulse delay of $\tau=7~\mu$s is shown in the inset of Fig.~\ref{Echo_decay}. For each echo signal, we extract the pulse envelope, fit it to the $y=y_0+A \cdot \textrm{sech}(\Delta\tau/w)^2$ (where $A$ is the amplitude, the $\Delta\tau$ is the detuning from the maximum of the echo envelope, $w$ is the width of the echo, and $y_0$ is the background noise level) 
and plot its intensity, $A^2$, as a function of the delay between excitation pulses $\tau$. 
The Figure~\ref{Echo_decay} displays the normalized echo decay measured at the magnetic field of 280~mT at cryostat temperatures of 0.012~K, 0.15~K and 0.4~K. The echo decays are non-exponential as the consequence of the spectral diffusion.

In order to reduce the heating of the sample due to excess of non-equilibrium phonons and to let the spins fully relax to their ground state, the echo sequence is repeated at an extremely slow rate of 0.5~Hz which is much slower than the optical relaxation $T_1\simeq 100$~Hz 

\section{Basic equations}
\label{sec:5}

To analyse the echo decays, we use the generalized formula proposed by B\"{o}ttger et al. \cite{Boettger2006}:
\begin{eqnarray}
I(t_{12},t_{23})& = & I_0  e^{-2t_{23}/T_1} \nonumber \\
& & e^{-4\pi t_{12} (\Gamma_0 + {1 \over 2} \Gamma_{SD}(Rt_{12}+1-e^{-Rt_{23}})) },
\label{eq_exp_fit}
\end{eqnarray}
where $t_{12}$ and $t_{23}$ are the first-to-second and second-to-third pulse delays respectively (for the 2PE, $t_{23} = 0$); $R$ is the characteristic relaxation rate; $\Gamma_0$ is the homogeneous linewidth; $\Gamma_{SD}$ is the spectral diffusion (SD) linewidth; $T_1$ is the relaxation time of the excited state.
The dephasing time $T_M$ is then derived as follows \cite{Boettger2006}:
\begin{eqnarray}
T_M& = & {2\Gamma_0 \over \Gamma_{SD} R}\left(-1+{\Gamma_{SD}R \over \pi \Gamma_0^2}\right).
\label{eq:Tm}
\end{eqnarray}

The dephasing time contains information about nearly all processes influencing the spins, and it is derived from homogeneous linewidth $\Gamma_0$, magnetic dipolar broadening, a.k.a. spectral diffusion $\Gamma_{SD}$, and characteristic relaxation rate $R$~\cite{Boettger2006}. All these parameters can be extracted by performing two-pulse (2PE) and three-pulse echo (3PE) experiments. 
The echo experiments with Er:YSO at millikelvin temperature range have already demonstrated substantial slowing down of the spin-lattice relaxation process even at weak magnetic fields, where the spin-lattice relaxation time attains time-scale of seconds $T_1^{SLR}\sim 1-10~$sec~\cite{Probst2013,Tkalcec2014}. 

Each source of decoherence in Eqs.\,(\ref{eq_exp_fit}-\ref{eq:Tm}) has a characteristic dependence on the magnetic field and temperature. 
The relaxation rate $R$ includes decoherence from spin-spin and spin-phonon interactions together, e.g. flip-flops, Raman and Orbach processes, direct process, i.e. processes directly involving the resonant spins.
In our thermal range, the relaxation rate $R$ is narrowed down to the flip-flop $R_{ff}$ and direct $R_d$ processes: $R= R_{d} + R_{ff}$. The flip-flop rate is given by the resonant spin flips \cite{Abragam1970}:
\begin{eqnarray}
R_{ff}=W_{ff}\textrm{tanh}\bigg({\hbar\omega \over k_B T_{spin}}\bigg)^2,
\end{eqnarray}
where $W_{ff}$ is the flip-flop coefficient, $\omega$ is the angular frequency of the transition between the spin Zeeman states, $T_{spin}$ is the temperature of the spin system~\cite{Kukharchyk2017}, $k_B$ is the Boltzmann constant. 

The direct process rate $R_d$ is generally composed of the direct and phonon bottleneck processes (PBN), which shows up at temperatures below 1 Kelvin~\cite{Abragam1970}:
\begin{eqnarray}
R_d = {1 \over \tau_{1d} + (1+b)\tau_{ph} }, 
\label{eq:Rd1}
\end{eqnarray}  
where $\tau_{1d}$ is the spin-phonon relaxation time, inverse of which is understood as the direct process in the absence of the bottleneck; $\tau_{ph}$ is the lifetime of the phonons set by the mean free path of a phonon; $b$ is the bottleneck coefficient. At low temperatures, $\tau_{ph}$ is known to be 500~ns \cite{Wild1998}, which also correlates with our estimate via heat capacity, $C_p$, and thermal conductivity, $\kappa$: $C_p r^2 / 4 \kappa\simeq 300$~ns. The estimate of the spin-phonon relaxation time  gives $\tau_{1d}\sim10^{-5}$~s~\cite{Abragam1970} for the temperatures below 1~K. 
The bottleneck coefficient $b$ is given by: 
\begin{eqnarray}
b={n_{at} \over \Sigma_{ph}} \textrm{tanh}\bigg({\hbar\omega \over k_B T_{spin}}\bigg)^2\sim 10^3,
\end{eqnarray}
where $n_{at}\simeq3.66\cdot10^{20}$~m$^{-3}$ is the density of Erbium ions; $\Sigma_{ph}$ is the density of phonons which are resonant to the inhomogeneous spin linewidth $\Delta\omega$~\cite{Probst2013,Probst2015}. $\Sigma_{ph}$ is estimated as~\cite{Abragam1970}
\begin{eqnarray}
\Sigma_{ph}={2\omega^2 \Delta\omega \over 2\pi \upsilon^3} = 1.86\cdot10^{17} m^{-3},
\end{eqnarray}
where $\upsilon$ is the speed of sound in the substrate. We thus obtain that the direct process rate $R_d$ is dominated by the bottleneck relaxation: $\tau_{1d} \ll \{(1+b)\tau_{ph}\simeq 0.5\cdot10^3 \}$, and the Eq.~\ref{eq:Rd1} is simplified to 
\begin{eqnarray}
R_d &=& {1 \over b\tau_{ph} } \nonumber\\
&=&{3 \mu_B^2\Delta\nu \over \hbar^2\upsilon^3 n_{at}\tau_{ph}}g^2B^2\textrm{coth}\bigg({\hbar\omega \over k_B T_{spin}}\bigg)^2.
\label{Rd_final}
\end{eqnarray} 
Thus, the complete equation for the relaxation rate $R$ reads as:
\begin{eqnarray}
R&=&W_{BN}\,g^2B^2\textrm{coth}\bigg({\hbar\omega \over k_B T_{spin}}\bigg)^2\\
&+&W_{ff}\,\textrm{tanh}\bigg({\hbar\omega \over k_B T_{spin}}\bigg)^2.
\label{R_fit}
\end{eqnarray}
Taking the speed of sound in Y$_2$SiO$_5$ equal to $\upsilon\simeq 4$~km/s and $\Delta\omega\simeq2\pi\cdot30$~MHz, we estimate the bottleneck coefficient $W_{BN} \simeq 2\pi\cdot6$~kHz\,T$^{-2}$. The flip-flop coefficient is estimated to be $W_{ff}\simeq 2\pi\cdot8$~kHz \cite{Kukharchyk2017,Abragam1970}.

The spectral diffusion reveals the contribution of the indirect flip-flop processes to the coherence and is typically described as \cite{Boettger2006}
	\begin{eqnarray}
	\Gamma_{SD} = \Gamma_{max}\textrm{sech}^2{g_{env}\mu_BB \over 2k_B T_{spin}},
	\label{eq_Gsd}
	\end{eqnarray}
	where $\Gamma_{max}$ is the FWHM frequency broadening resulting from the magnetic dipole-dipole interactions. Spectral diffusion can result from flip-flops of Erbium spins of Site 1 and of the flip-flops of nuclear Yttrium spins. Spectral diffusion rate by Yttrium spins has already been discussed by Böttger~et.~al.~\cite{Boettger2006}. Estimated to be $\simeq1$~Hz, spectral diffusion by Yttrium ions can be neglected in current study. Spectral diffusion by Site~1 Erbium is  estimated to be $\Gamma_{max}\simeq 2\pi\cdot60$~kHz for our crystal orientation~\cite{Boettger2006}.

	The homogeneous linewidth $\Gamma_0$ contains contributions from dynamic broadening mechanisms which occur faster than the experimental time-scale, such as the homogeneous linewidth $\Gamma_h = 1/ \pi T_2$ given by the lifetime broadening and single-ion linewidth, optical broadening due to the spin flips of the ground state $\Delta \Gamma_h$, and fast spectral diffusion $\Gamma_{SDh}$ resulting from indirect spin flip-slops:
	\begin{eqnarray}
	\Gamma_0 = \Gamma_h + \Delta \Gamma_h + \Gamma_{SDh},
	\label{G0_eq}
	\end{eqnarray}
	where $\Delta \Gamma_h$ is given by \cite{Boettger2006}
	\begin{eqnarray}
	\Delta \Gamma_h = {R \over 4\pi}e^{-g_{g}\mu_BB \over 2k_B T}\textrm{sech}{g_{env}\mu_BB \over 2k_B T_{spin}};
	\label{eq:dGh}
	\end{eqnarray}
	$\Gamma_{SDh}$ is the contribution from the spectral diffusion, which cannot be resolved as part of $\Gamma_{SD}$ in the Eq.~\ref{eq_exp_fit}, and is thus described by the same dependence on the magnetic field and temperature as the spectral diffusion $\Gamma_{SD}$:
	\begin{eqnarray}
	\Gamma_{SDh} = \Gamma_{maxh} \textrm{sech}^2{g_{env}\mu_BB \over 2k_B T_{spin}}.
	\end{eqnarray}

Measured via stimulated echo decay, the value of the relaxation time T$_1$ is very sensitive to the decoherence and spectral diffusion occurring during the both $t_{12}$ and $t_{23}$ delays. The model in Eq.~(\ref{eq_exp_fit}) developed by B\"{o}ttger et al. \cite{Boettger2006} aims to include decoherence occurring during the both delays
	given that there is only one source of SD.  It, however, does not include any contributions from the non-equilibrium phonons evolving during $t_{12}$ or more than one source of SD.

 Temperature measured from the sensor in the cryostat is typically different from actual temperature of the spin system, $T_{spin}$. The resonant excitation of the ions first excites the spin-system out of the thermal equilibrium which is followed by the fluorescence and population of spin-flipped states.
Upon further relaxation, part of the energy absorbed by the spins is transferred into phonons. This in turn rises local temperature of the crystal. Also, it is possible that part of the optical excitation can be directly absorbed in the volume of the crystal which is overlapping with the laser beam volume. Thermal properties of the crystal, i.e. thermal capacity and conductance, are governing time over  which the system comes to the thermal equilibrium. It is not possible to measure such evolution of local temperature with a sensor. However, it is possible to extract effective temperature of the spin-system $T_{spin}$ from the data.  At higher temperatures, $T_{spin}$ equals to the temperature of the thermal sensor, however, at lower temperatures, $T_{spin}$ is saturated at a minimal thermal point $T_{min}$ , s.f.~\cite{Kukharchyk2018}. We find that it can be well described by the equation
	\begin{eqnarray}
	T_{eff}=T_{min}\bigg(1+\bigg({T \over T_{min}}\bigg)^2\bigg)^{1/2},
	\end{eqnarray}
	where $T_{min}$ is the minimal attainable temperature, and $T$ is the temperature measured by sensor. During the field dependence measurements, the temperature on the sensor of the cryostat equals 12~mK which is smaller than the minimal temperature attainable by the spin system,  $T \textless T_{min}$, therefore $T_{spin} \simeq T_{min}$.

\section{VI. Experimental results}
\begin{figure}[ht!]
		\includegraphics[width=\columnwidth]{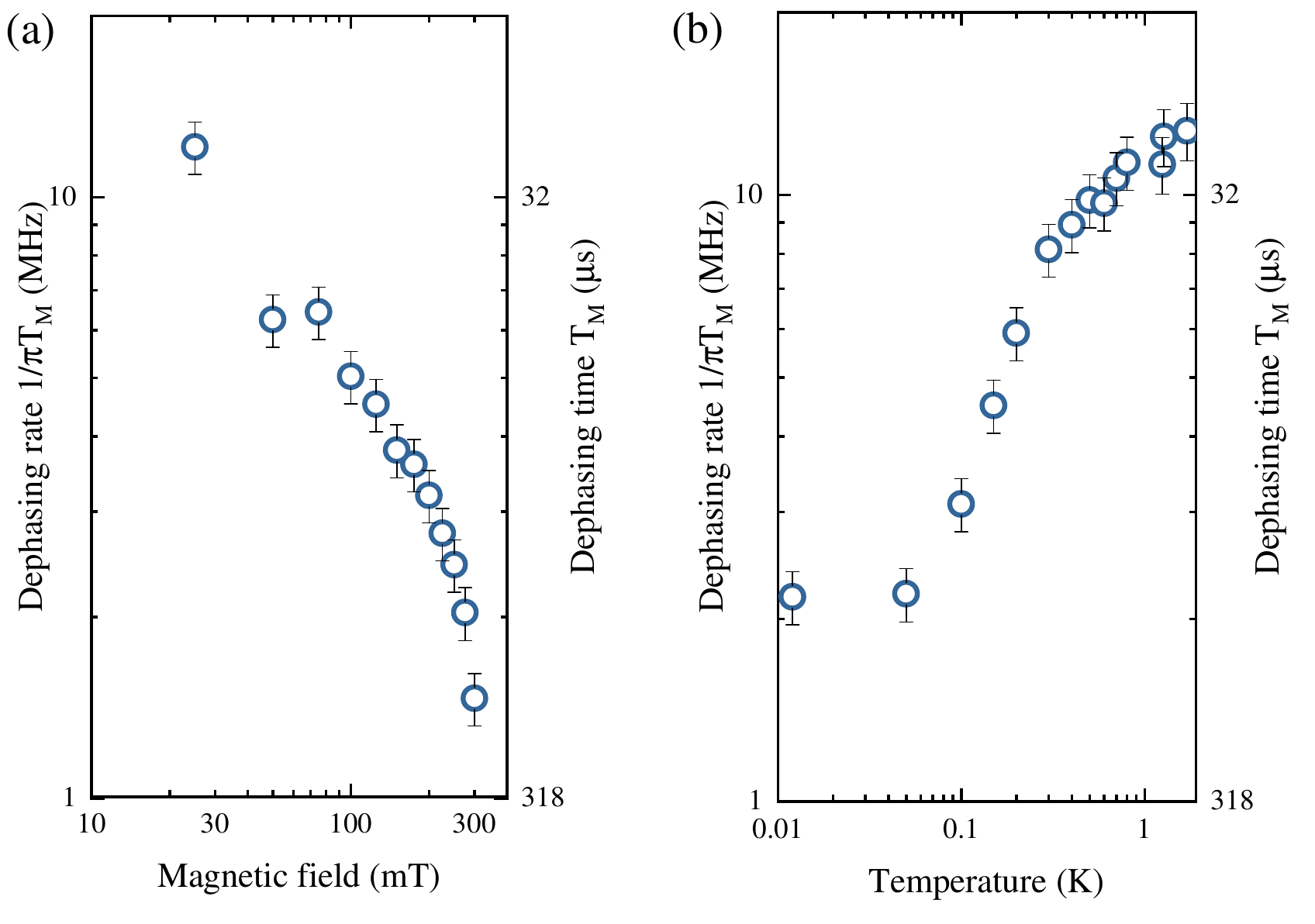}
		\caption{(Color online) (a) Decoherence rate $1 /\pi T_M$ as a function of the magnetic field at $12~$mK. (b).  Decoherence rate $1 /\pi T_M$ as a function of temperature at the magnetic field of $280~$mT.}
	\label{Tm_dep}
\end{figure}

\paragraph{Dephasing time}
The values of $T_M$ at different temperatures and magnetic fields are presented in Fig.~\ref{Tm_dep}. 
With increase of the magnetic field up to 300~mT, we observe the increase of the dephasing time by one order of magnitude: from 27~$\mu$s at 30~mT to 217~$\mu$s at 300~mT. Similarly, $T_M$ increases by one order of magnitude with the decrease of temperature. The most dramatic increase of $T_M$ happens below 500~mK till the minimal temperature is reached: from $(39\pm2)~\mu$s to $(146\pm7)~\mu$s. This is due to the fast polarization of the spins when thermal energy drops below 10~GHz \cite{Takahashi2008}. Above 900~mK, $T_M$ remains nearly $\simeq 30~\mu s$. Below 100~mK, $T_M$ saturates which suggests that the minimal attainable temperature is $T_{min} \sim (50 - 100)$~mK, see Fig.~\ref{Tm_dep}(b).

\begin{figure}[ht!]
	
	\includegraphics[width=\columnwidth]{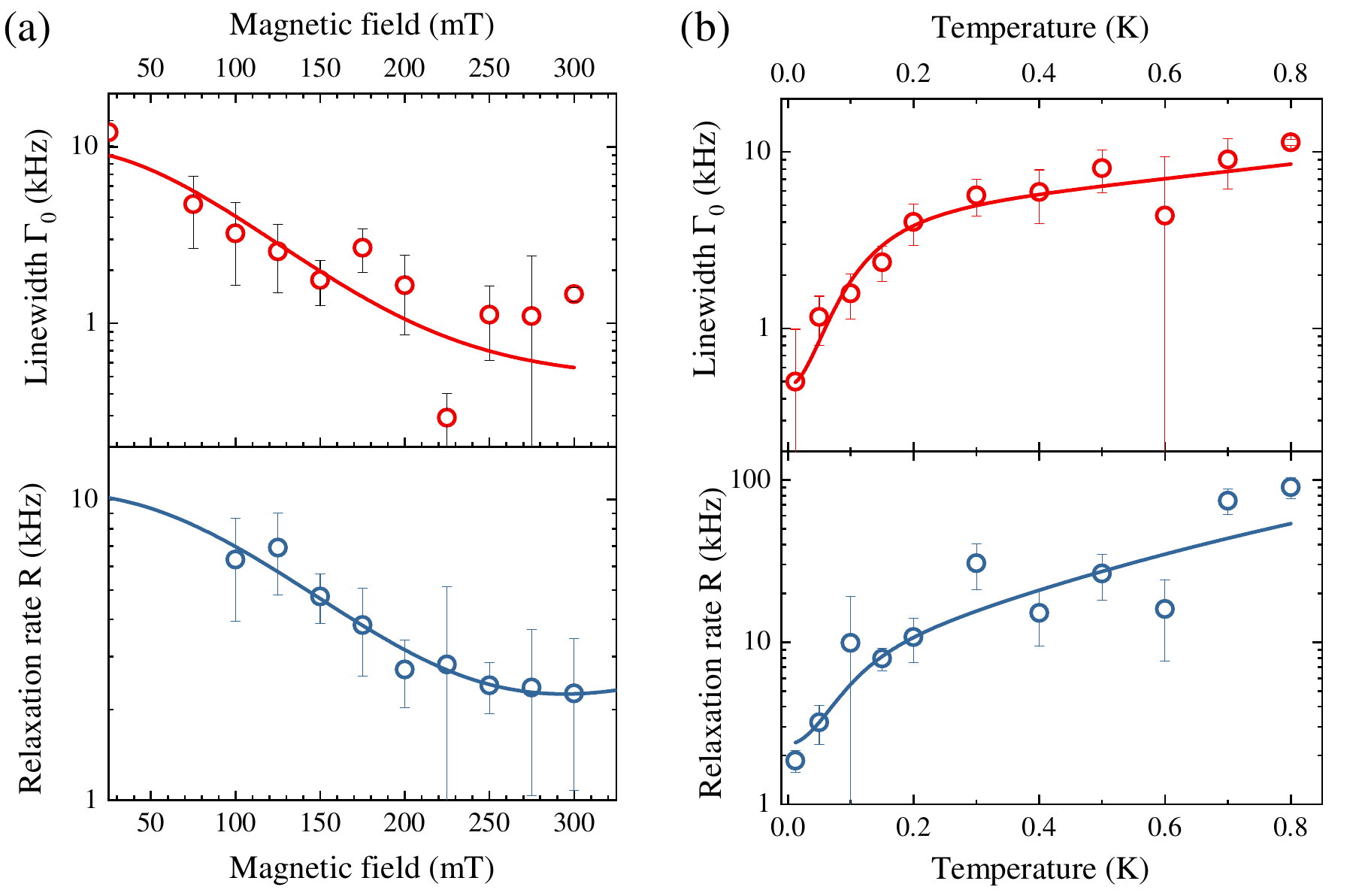}
	
	\caption{(Color online) (a) Dependence of the homogeneous linewidth $\Gamma_0$ and relaxation rate $R$ on magnetic field. (b) Dependence of the homogeneous linewidth $\Gamma_0$ and relaxation rate $R$ on temperature at the magnetic field of 280~mT. The red line represents fit of the data to Eq.~\ref{R_fit} and Eq.~\ref{G0_eq}.}
	\label{RG0_dep}
\end{figure}

\paragraph{Relaxation rate R}
To extract the decoherence processes, we employ the model explained in Section~\ref{sec:5}, and the obtained parameters are summarized in Table~\ref{table}. 
Values of the relaxation rate $R$ and their fit to the Eq.~(\ref{R_fit}) are shown in Fig.~\ref{RG0_dep}. 
Both magnetic field and temperature dependencies of the relaxation rate converge to same parameters.
The flip-flop rate equals to $W_{ff}\simeq2\pi\cdot(10.0\pm0.2)$~kHz, which is similar to the analytically estimated value of $\simeq2\pi\cdot8$~kHz \cite{Abragam1970}.
The phonon-bottleneck rate equals $W_{BN}\simeq2\pi\cdot(7.5\pm0.2)$~kHz\,T$^{-2}$ and agrees with analytical $W_{BN}\simeq2\pi\cdot6$~kHz\,T$^{-2}$. 
The minimal temperature attained by the coherently excited spin-system is $T_{min}\simeq2\pi\cdot(74\pm9)$~mK which fits to the observed saturation of the $T_M$ below 100~mK, see Fig.~\ref{Tm_dep}.

\paragraph{Homogeneous linewidth}
The homogeneous linewidth $\Gamma_0$ shows dependence on the magnetic field and temperature, see Fig.~\ref{RG0_dep}. 
To fit $\Gamma_0$ to Eq.~\ref{G0_eq}, we define $\Delta\Gamma_h$ by using the relaxation rate with parameters from the fit, and insert it into the Eq.~\ref{eq:dGh}. The resulting
fit of Eq.~\ref{G0_eq} to the experimental data is shown in Fig.~\ref{RG0_dep}. The minimal temperature measured through the $\Gamma_0$ is the same as for the relaxation rate $R$, $T_{min}\simeq 74\pm9$~mK.
$\Gamma_h$ equals $\simeq2\pi\cdot0.3$~kHz, which is smaller than $\Gamma_0$ vales obtained by Böttger~et.~al.~\cite{Boettger2006}: $2\pi\cdot$22~kHz at 4.2~K and $2\pi\cdot$1.3~kHz at 1.6~K for the erbium concentration of 0.005 at.$\%$, which is expected due to lower temperature of the spin-system and the thermal bath in our experiment.
The amplitude of spectral diffusion extracted from $\Gamma_0$ is different in case of dependence on the magnetic field, $\Gamma_{maxh}\simeq 2\pi\cdot(11.4\pm2.6)$~kHz, and in case of temperature dependence, $\Gamma_{maxh}\simeq 2\pi\cdot(3.3\pm0.5)$~kHz. 
In both magnetic field and temperature dependences, major contribution to the $\Gamma_0$ is due to spectral diffusion $\Gamma_{SDh}$, up to $\sim90\%$ of the value of $\Gamma_0$.


\begin{figure}[ht!]
	
	\includegraphics[width=\columnwidth]{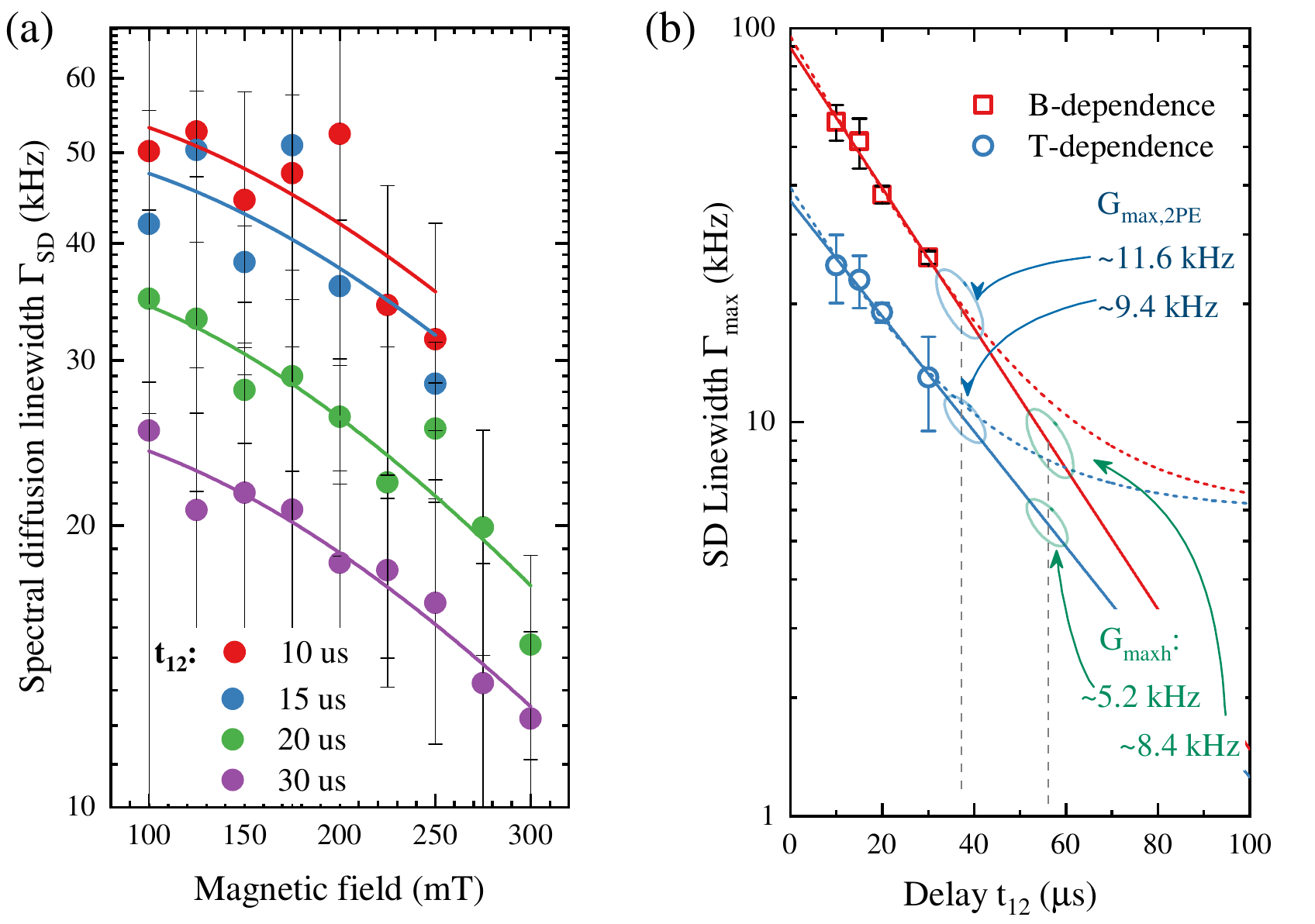}
	
	\caption{(Color online) (a) Spectral Diffusion linewidth $\Gamma_{SD}$ as a function of the magnetic field for different $t_{12}$ values. Lines show the fits of the data to Eq.~\ref{eq_Gsd}. (b) $\Gamma_{max}$ values obtained from the fit at different $t_{12}$. Red and black lines show the exponential fit of the $\Gamma_{max}$ dependence on the $t_{12}$. Possible position of the $\Gamma_{max2PE}$ and $\Gamma_{maxh}$ on the exponential decay is shown with the ovals, the dashed line guides to a corresponding $t_{12}$ delay.}
	\label{Gsd_dep}
\end{figure}

\paragraph{Spectral diffusion}
The discrepancy between $\Gamma_{SDh}$ values in magnetic field and temperature dependencies is similar to the discrepancy in spectral diffusion $\Gamma_{SD}$ values, which is derived directly by the Eq.~\ref{eq_exp_fit}. Additionally,
 the obtained SD amplitude $\Gamma_{max}$ depends on the delay $t_{12}$ in 3PE experiments, which has not been observed in the experimental investigations of Er:Y$_2$SiO$_5$ before~\cite{Boettger2006}.
Dependence of the spectral diffusion on the magnetic field and $t_{12}$-delay
is shown in Fig.\ref{Gsd_dep}(a). The minimal attainable temperature $T_{min}$ converges to $\simeq 170$~mK for all datasets. 
$\Gamma_{max}$ decreases from $2\pi\cdot$58~kHz ($2\pi\cdot22$~kHz) for $t_{12} = 10~\mu$s to $2\pi\cdot$26~kHz ($2\pi\cdot13$~kHz) for $t_{12} = 30~\mu$s for magnetic field (temperature) dependence, see Tab.~\ref{table} and Fig.~\ref{Gsd_dep}(b). 
The $\Gamma_{max}$ values obtained from the temperature dependence are $\sim2$ times smaller than $\Gamma_{max}$ values obtained from the field dependence. 
We relate such dependence of the $\Gamma_{max}$ on $t_{12}$ and on the field/temperature as well as higher $T_{min}$ to the dynamics of non-equilibrium phonons (NQP) and to the dependence of SD on the dynamics of NQP, which we discuss in detail in the next section.
 
The $\Gamma_{max}$ values are nicely following an exponential dependence on $t_{12}$, see Fig.~\ref{Gsd_dep}\,(b), with the characteristic time $\tau_{t12}\simeq~25~\mu$s, which holds for both temperature and B-field dependencies.
From the exponential dependence of $\Gamma_{max}$ on $t_{12}$, we derive the maximal amplitude of $\Gamma_{max}(t_{12}\rightarrow0)\simeq 2\pi\cdot90$~kHz for the B-field dependence and $\Gamma_{max}(t_{12}\rightarrow0)\simeq 2\pi\cdot35$~kHz for the temperature dependence, which are within 50$\%$ variation of the estimated value $\Gamma_{max}\simeq2\pi\cdot60$~kHz.

	





In the 2PE experiment, we extract the amplide of spectral diffusion from the product $\Gamma_{SD}R$ by using the relaxation rate $R$ parameters extrated earlier in the 3PE experiment. Thus, amplitudes of spectral diffusion derived from magnetic field dependence, $\Gamma_{max}\simeq2\pi\cdot12$~kHz, and from temperature dependence, $\Gamma_{max}\simeq2\pi\cdot9$~kHz, also vary by a factor of $\simeq$2. These values are smaller that in the 3PE experiment, and if placed onto the exponential dependence on $t_{12}$, both values correspond to the  $t_{12}\simeq37~\mu$s in the 3PE experiment. 
Value of T$_{min}$ is in agreement for both 2PE and 3PE experiments and equals $\simeq(163\pm79)$~mK in the 2PE experiment.


\begin{figure}[ht!]
	\includegraphics[width=\columnwidth]{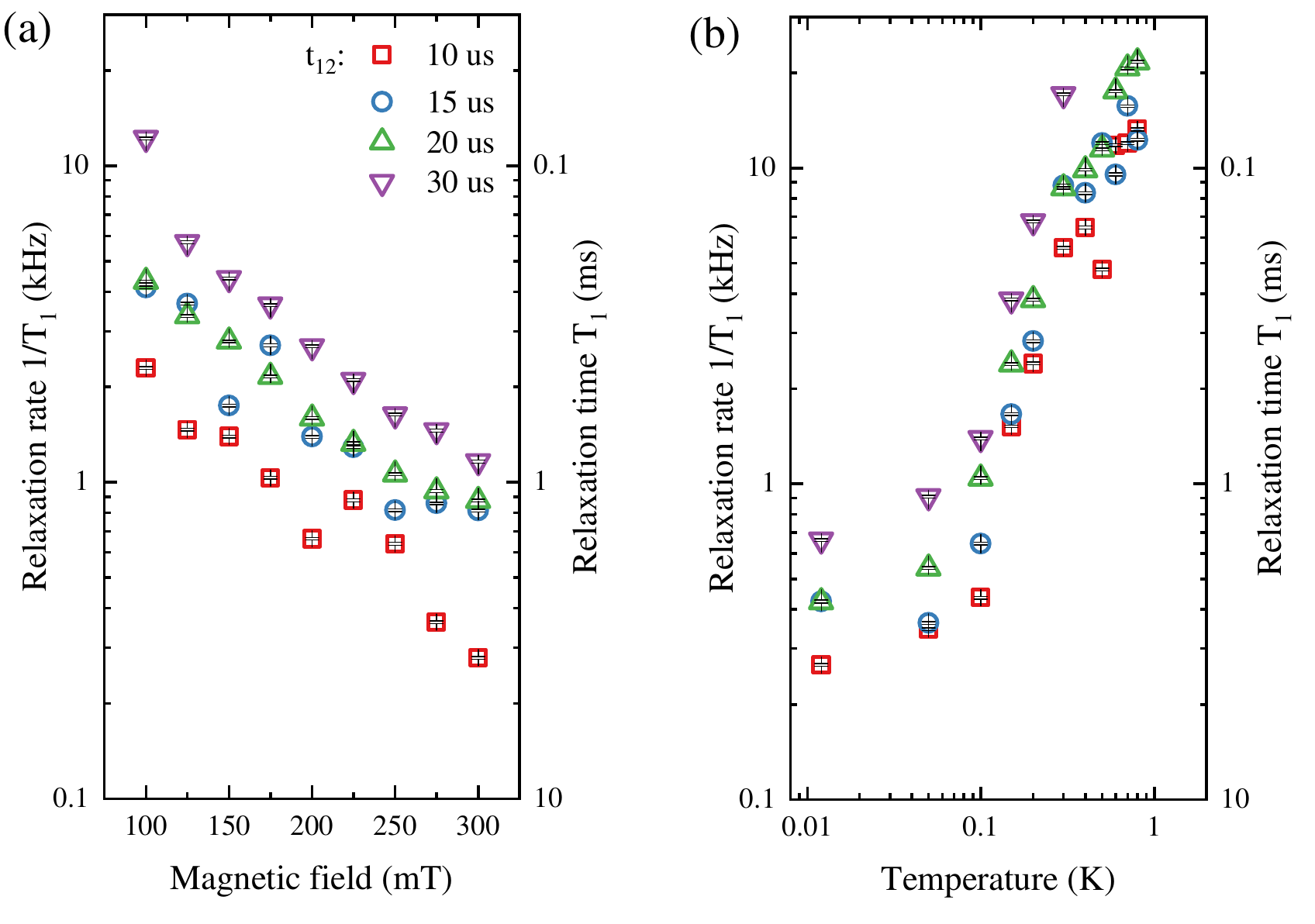}
	\caption{(Color online) (a)~SLR rate $1 /\pi T_1$ as a function of the magnetic field at $12~$mK for different values of t$_{12}$. (b)~SLR rate $1 /\pi T_1$ as a function of temperature at the magnetic field of $280~$mT for different values of t$_{12}$. Legend holds for both (a) and (b)}
	\label{T1_dep}
\end{figure}

\paragraph{Relaxation time T$_1$}
The relaxation time $T_1$ is the characteristic optical relaxation time. It is derived by fitting the 3PE decays to the Eq.~\ref{eq_exp_fit}.
The relaxation time $T_1$ increases with increase of the magnetic field and reaches $\sim$3500~$\mu s$ at 300~mT for $t_{12} = 10~\mu$s, see Fig.~\ref{T1_dep}~(a), which close to the condition of the full polarization of spins. At low magnetic fields, i.e. 30~mT - 50~mT, $T_1 \simeq T_M \simeq 30~\mu s$. 
At 12~mK, we see that $T_1$ depends on the delay $t_{12}$, similarly to the spectral diffusion described in the section above. In the temperature dependence, see Fig.~\ref{T1_dep}, we see that this dependence on $t_{12}$ is much less relevant at temperatures above 0.5~K.
Thus below 0.5~K, the rotating $\pi/2$ pulses thermally influence the decoherence mechanisms, via creation of NQP with dynamics on the timescales of the experiment.

\begin{table*}[t]
\centering
\caption{Derived experimental and theoretical values}
\begin{tabular}{ |p{2.3cm}||p{1cm}|p{2cm}|p{2.7cm}|p{2cm}| p{3.8cm}| p{2cm}| }

	\hline
	& &  $\sfrac{1}{2\pi}W_{ff}$, kHz&  \sfrac{1}{$2\pi$}$W_{BN}$, kHz$\cdot$T$^{-2}$ & $T_{min}$, mK & \sfrac{1}{$2\pi$}$\Gamma_{max}$, kHz &\sfrac{1}{$2\pi$}$\Gamma_h$, Hz\\
	\hline
	Rate R & 3PE & $10.0\pm0.2$  & $7.5\pm0.2$  & $74\pm9$ &   - & -\\
	$\Gamma_{SD}$, $t_{12} \rightarrow 0$  & 3PE & - & - & $170\pm37$ & $ 90~/~35^{*}$& -\\
	$\Gamma_{SD}$ & 2PE & \textit{fixed at} $10$  & \textit{fixed at} $7.5$ & $163\pm79$ & $11.6\pm2.3~/~9.4\pm0.5^{*}$& -\\
	$\Gamma_0$ & 2PE & \textit{fixed at} $10$  & \textit{fixed at} $7.5$ & $74\pm9$ & $8.4\pm1.6~/~5.2\pm0.5^{*}$& 300\\
	\hline
	Our numerical estimation & - & $\sim8$& $\sim5.7$ & - & $\sim60$ & $\sim100^s$ \\
	\hline
	Taken from \cite{Boettger2006} & - & - & - & $1.6\cdot10^3$ & 820 & 1300\\
	\hline
\end{tabular}
\\
$^{*}$ Values differ for magnetic field and temperature dependences. Presented as\\ ( value from magnetic field dependence) / (value from temperature dependence).
\label{table}
\end{table*}

\section{Discussion}
The dependence of the spectral diffusion amplitude, $\Gamma_{max}$, and optical relaxation time, $T_1$, on the pulse delay $t_{12}$ allows to suggest that the model in Eq.~(\ref{eq_exp_fit}) is not capable of fully describing all the decoherence phenomena at our experimental conditions. As fact, contributions of the actual dynamics of the NQP and the spectral diffusion in the presence of NQP need to be developed. We further discuss the qualitative picture of the processes occurring in the photon echo experiments at millikelvin temperatures and possibly leading to the declination from the model.

\begin{figure}[ht!]
		\includegraphics[width=\columnwidth]{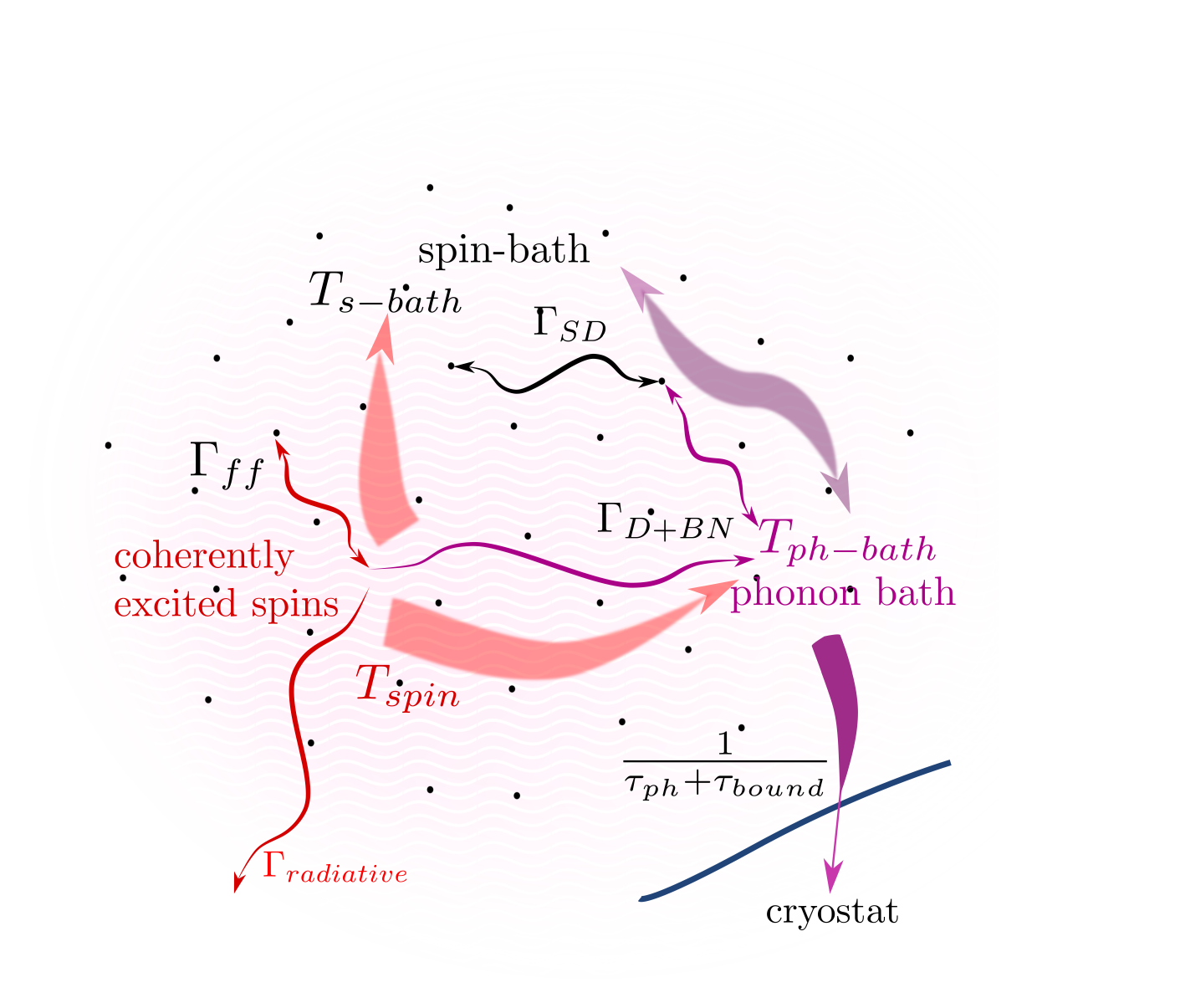}
		\caption{(Color online) Illustration of decoherence and relaxation prosesses taking place in the crystal at our experimental conditions}
	\label{fig:relax_ill}
\end{figure}

Let us consider four sub-systems: an ensemble of coherently excited erbium ions, the spin bath of erbium ions, the phonon bath of the host crystal and the cryostat thermal bath Fig.~\ref{fig:relax_ill}. Before the first pulse arrives, all these sub-systems are at thermal equilibrium, and a specific number of thermal phonons exists in the crystal, which is given by the Boltzmann distribution. Upon the arrival of the first pulse, part of the pulse energy is absorbed by the atoms, another part - is  absorbed by the crystals and is directly converted into the phononic excitation of the host crystal, and the rest of it leaves the crystal un-absorbed. Resulting from the excitation of the crystal by the pulse and by the non-radiative emission from the ions, a number of non-equilibrium phonons (NQP) is created. These phonons are associated with a locally increasing temperature of the crystal that exceeds the thermal equilibrium temperature, which however is not observed with the thermal sensor of the cryostat. These phonons travel through the crystal, participate in collisions with other phonons, and leave the crystal through the boundary interface.

The influence of non-equilibrium phonons (NQP) created by a laser pulse has been already considered in different crystals including RE-doped Y$_2$SiO$_5$~\cite{Macfarlane1985,Bai1992,Altner1996}. 
The number of created NQP depends on the excitation energy, resonant phonon frequencies, presence of the other phonon effects. NQP lead to the faster disappearance of the echo envelope and thus to the shorter measured coherence and relaxation times, which we observe in our data controversially: the relaxation time $T_1$ is longer for a shorter delay $t_{12}$, similarly to the SD. The NQP are tightly linked to the PBN effect, which appears when the NQP cannot leave the excited volume fast enough and the amount of energy stored in the ions and to be released into phonons exceeds the number of resonant phonon frequencies.

Let us consider specific thermal properties of the atoms and the substrate. The heat capacity of the YSO crystal can be estimated from the Debye model\cite{Kittel2005,Pobell2007} or taken from tables\cite{Pobell2007}, and is $\sim3$ times smaller than that of the SiO$_2$ and equals $c_{YSO}\simeq0.22\cdot T^3 {J \over m^3 K}$. We assume that the thermal conductivity of the YSO below 1~K is in the order of that of the SiO$_2$, which is $\kappa_{YSO}\simeq0.02\,T^2 {W \over m K}$\cite{Pobell2007}. 

The heat capacity of a spin system at ultralow temperatures is described by the well-known Shottky anomaly equation\cite{Sears1964,Tucker1965}: $c_{spin} = (\Delta E/k_BT)^2 e^{(-\Delta E/k_BT)}/(1+e^{(-\Delta E/k_BT)})^2$, where $\Delta E$ is the Zeeman splitting energy of the spin system. The thermal capacity $c_{spin}$ has a characteristic peak exactly in our working field-temperature range.
The Shottky anomaly in $c_{spin}$ occurs between 50~mK and 500~mK for the magnetic field range $(50 - 300)$~mT resulting into $c_{spin}\simeq c_{YSO}\cdot 10^3$. 
This means that the spin system can store 1000 more thermal energy than the crystal the same range of temperature and fields.  Taking into account the echo efficiency, see the inset in Fig.~\ref{Echo_decay}, a large part of this energy will be non-radiatively transmitted into NQP. Only if the NQP lifetime in the crystal is much shorter than the coherence time and that the thermal coupling to the cryostat is good, phonons will leave the crystal fast with no influence on the measured echo signal. Below 1~K, resistance of thermal boundary usually leads to the reduction of the heat transfer through the sample-cryostat contact, which is known as Kapitza resistance \cite{Kapitza1941,Liberadzka2019,Kukharchyk2017}. The NQPs can not leave the crystal fast enough and block the paths for the spins to relax, moreover they non-coherently re-excite other spins and thus increase the effect of spectral diffusion. As the result, the phonon bottleneck effect appears, and the dynamics of the non-equilibrium phonons enters the dynamics of the spin system itself.

The energy transfer between the sub-systems is schematically depicted in Fig.~\ref{fig:relax_ill}. The non-radiative relaxation of the coherently excited spins transfers energy to the spin bath and to the phonon bath. Phonon bath in the presence of the phonon bottleneck effect transfers part of the energy to the spin bath as well. This leads to the increase of the temperature of the spin bath, which we observe as high effective temperature of the spectral diffusion, which suggests that the actual temperature of the local phonon bath is also higher than the temperature of the spin bath. We do not posses a good model to simulate it, it is however important to outline that the temperature dependencies of flip-flops, direct process and phonon bottleneck imply that temperatures of the local sub-systems are equal. This should be expected at higher measurement temperatures, above 1K, when the change of the temperature due to the excitation would be neglectable, but not at millikelvin temperatures.

We do not know how large the thermal resistance of the boundary is in our experiment. Based on the observation of the phonon bottleneck effect, we assume it to be rather large. Degradation of the thermal contact~\cite{Liberadzka2019} and high density of the NQP lead to the dependence of the measured spectral diffusion $\Gamma_{SD}$ and relaxation time $T_1$ on the $t_{12}$ delay which is not included in the current general model. We also see that the set of the t$_{12}$ delays is in the order of pure thermal relaxation in our crystal witout consideration of the thermal contact. 
$T_1$ is known to be sensitive to the final conditions of the spin-crystal system by the moment of arrival of the second pulse. 
So for the shorter $t_{12}$ delays, larger number of the NQP is left inside the crystal, therefore, $T_1$ values are smaller.
For the longer $t_{12}$ delay, effect of the first $\pi$/2 pulse on $T_1$ is reduced and $T_1$ values are thus larger.

Spectral diffusion linewidth  $\Gamma_{max}$  contains a cumulative effect from NQP. $\Gamma_{max}$ is larger for the shorter $t_{12}$ delay and decreases in an exponential way. We can attribute the characteristic time $\tau_{t12}$ of this exponential dependence to the lifetime of the NQP in the crystal~\cite{Bai1992}. It then explains, why we observe such a drastic influence of the $t_{12}$ delay on experimental result: relaxation timescale of the NQP is in the order of the $t_{12}$ delay between the pulses. Extension of the phonon lifetime by nearly two orders (compare,  $\tau_{t12}\simeq25~\mu$s and $\tau_{ph}\simeq0.5~\mu$s) is due to the strong phonon bottleneck effect and high thermal boundary resistance.
We can assess the contribution from the NQP to the relaxation as follows \cite{Graf1998}:
\begin{eqnarray}
\Gamma_{NQP}&=&{\sigma_0 \upsilon \Sigma_{ph} \over \pi}\simeq 2\pi\cdot 10~kHz,
\end{eqnarray} 
where $\sigma_0 \simeq 10~nm^2$\cite{Graf1998} is the phonon cross-section. $\Gamma_{NQP}$ is in fact nothing else than the bottleneck coefficient $1/\tau b$, which brings it to the order of the bottleneck coefficient.

\section{Conclusions}

In conclusion, we have presented a detailed coherent optical spectroscopy of mono-isotopic $^{167}$Er:YSO crystal at moderate fields and temperatures below 1~K.
In the echo experiments, we have studied the decoherence properties at low magnetic fields and millikelvin temperatures. The main sources of decoherence are the spectral diffusion, non-equilibrium phonons and phonon bottleneck effect. 
Spectral diffusion linewidth and relaxation rate are particular sensitive to the dynamics of the non-equilibrium phonons, which occurs in the timescale of the experiment. 
As the limiting factor, we find the thermal properties of the crystal itself: heat capacity is 1000-times smaller than that of the spin-system. Deterioration of the thermal interface below 0.5~K leads to the higher number of active phonon modes during the echo measurement, which limit the maximal achievable coherence time. 
It is thus not possible to achieve an increase in optical coherence by solely cooling a spin-doped crystal down to millikelvin temperatures. We thus can conclude that in bulk samples it is much easier to freeze the spins by applying the magnetic field rather than by cooling down.

\section{Acknowledgement}

This work is supported by the Saarland University, Land of Saarland and DFG through the grant INST 256/415-1, BU 2510/2-1. A.K. acknowledges support by the government assignment for FRC Kazan Scientific Center of RAS.

\bibliographystyle{abbrv}
\bibliography{ErYSO}

\end{document}